\documentclass[twocolumn,showpacs,preprintnumbers,amsmath,amssymb]{revtex4}

\usepackage{dcolumn}
\usepackage{bm}

\newcommand{\p}[1]{(\ref{#1})}

\newcommand{\A}{\mathfrak{A}}

\begin{document}


\title{The Spinning Particles as a Nonlinear Realizations of the
Superworldline Reparametrization Invariance}

\author{J. Malinsky}%
\altaffiliation[Also at ]{Department of Biomathematics,
 Box 1213, Mount Sinai Medical Center, New York, New York 10028}
 \author{V.P. Akulov}%
 \email{akulov@juno.com}
\author{N. Abdellatif}
\affiliation{%
Department of Physics and Technology,
Bronx Community College of the City University of New York,\\
Bronx, New York 10453
}%

\author{A. Pashnev}
 \email{pashnev@thsun1.jinr.ru}
\affiliation{
Bogoliubov Laboratory of Theoretical Physics, JINR \\
 Dubna, 141980, Russia
}%

\date{\today}

\begin{abstract}
The superdiffeomorphisms invariant description
of $N$ - extended spinning particle is constructed in the framework of
nonlinear realizations approach. The action is universal
for all values of $N$ and describes the time evolution of $D+2$ different
group elements of the superdiffeomorphisms group of the $(1,N)$ superspace.
The form of this action
coincides  with the one-dimensional
version of the gravity action,  analogous to Trautman's one.
\end{abstract}


\maketitle

\section{Introduction}
As is well known there exist several equivalent formulations of the
massless relativistic particles. The second order and first order
formalisms are examples of them.
One more example is the
conformally invariant description \cite{M}, which starts from
$D+2$ dimensional spacetime.
The existence of alternative approaches always sheds some new light
on the nature of the physical system. In particular,
as opposite to the conventional description of relativistic massless
particles, when the particle coordinates and einbein being the fundamental
variables play
essentially different roles, in the conformally invariant description
both of these types of variables are constructed from the more
fundamental
ones\cite{M,S}. The geometrical nature of these
initial variables
has been understood\cite{P1}-\cite{P2} in the framework of
the nonlinear realizations approach and their connection  with the
dilaton(s) of the reparametrization symmetry of the worldline of the particle
was established.

When it applied to different physical systems,
the method of nonlinear realizations\cite{CWZ}-\cite{V}  allows to understand
the geometrical meaning of the basic variables of the system
and construct the corresponding Lagrangians following some standard
procedure.

As was shown in \cite{BO}, gravity can be treated as a nonlinear
realization of the four dimensional diffeomorphisms group.
The consideration was based on the fact that infinite dimensional
diffeomorphisms group in four dimensional space can be represented as the
closure of two finite dimensional groups - conformal and affine ones \cite{OP}.
As a consequence of such representation of the diffeomorphisms group,
the basic field in this consideration was the symmetric tensor
field of the second rank - the metric field $g_{m,n}$, which
corresponds to symmetric generators of the affine group.
The generalization of this approach to the case of superspace was
given in \cite{IN}.

Alternatively, one can consider nonlinear realization of the {\it whole}
infinite dimensional diffeomorphisms group of the arbitrary
(super)space. Among the coordinates parameterizing the group element
(coset space)
in such realization there exist
usual coordinates of the (super)space.
The vielbeins and connections are  represented
by the other coordinates of the coset space. This approach
 in the case of $D$- dimensional bosonic space-time naturally leads to
 the Trautman's\cite{T} description of the gravity in terms of vielbeins
 and connection\cite{P_0}.
 In particular case of $D=1$ the same Lagrangian reproduces the
 conformally invariant description of the
massless relativistic particles\cite{PK}.
Though this approach was not applied to the $D$- dimensional
 superspace (to describe the supergravity),
its application to the $(1,N)$ superspace with one bosonic and
$N$ Grassmann coordinates $(\tau,\vartheta^a), (a=1,2,  \cdots N)$
 (spinning particles)
reveals many features of the bosonic case.

The einbein and its superpartners, as well as space-time
coordinates along with their own superpartners
describing the spin of the particle,
are connected to some parameters (dilaton and its superpartners)
parameterizing the superdiffeomorphisms group of the proper-time
superspace $(1,N)$. As it was shown\cite{P2}, there exist problems in
constructing the action in the superspace approach starting from
the case $N=3$. On the other hand, as it was claimed,
the component approach, developed in the paper\cite{P2} for $N=1$,
is applicable for an arbitrary value of $N$.

In the present paper we construct the conformally invariant
form\cite{M,S}
of the action for $N$ - extended spinning
particle\cite{GT,HPPT} in terms of geometrical quantities
of the reparametrization group of the proper-time
superspace $(1,N)$. The action is invariant
with respect to the general $(1,N)$ group of
 superdiffeomorphisms
and coincides with the conformally invariant
action\cite{M,S} after some gauge fixing.

\section{Nonlinear Realization of the Reparametrizations
in the $(1,N)$ Superspace}
With the help of the coordinate representation of the generators
in an auxiliary $(1,N)$ superspace $(s,\eta^a)$
\begin{eqnarray}                \label{rep}
{P^{\underbrace{\mbox{\scriptsize 0...0}}_{\mbox{\em\scriptsize m}}
a_1,a_2,...,a_n}}_0&=&
is^m\eta^{a_1}\eta^{a_2}... \eta^{a_n}\frac{\partial}
{\partial s},\\\nonumber
{P^{\underbrace{\mbox{\scriptsize 0...0}}_{\mbox{\em\scriptsize m}}a_1,a_2,...,a_n}}_a&=&
is^m\eta^{a_1}\eta^{a_2}... \eta^{a_n}\frac{\partial}
{\partial \eta^a},\quad  n\leq N.
\end{eqnarray}
one can calculate
the corresponding algebra of diffeomorphisms and forget
about this representation. In what follows these generators
will be considered as abstract ones. Using them one can
write down the {\em group} elements of the diffeomorphisms
{\em group}. Some parameters in the representation of this
group elements transform under the left multiplication
as the coordinates of the $(1,N)$ superspace. They
actually describe the proper-time superspace of the $N$ - extended
spinning particles.

To describe the $D$ - dimensional spinning particle in a conformally
invariant way we
introduce $(D+2)$  group elements in the following form $({\cal I}=
0,1,...,D+1)$:
\begin{eqnarray}\label{group}
G_{\cal I} &=& e^{{\rm i}\tau P_0}
 e^{{\rm i}\vartheta^a(\tau) P_a} \nonumber\\
&&  e^{{\rm i}\rho_a P^a_0}
e^{{\rm i}{\cal C}_{ab} P^{ab}_0}
e^{{\rm i}\chi_a (P^{0a}_0+{\rm i}P^0_a)}
e^{{\rm i}\zeta_{ab}^c P^{ab}_c}
e^{{\rm i}\A_a^b P^{0a}_b}
e^{{\rm i}\kappa^a P^{00}_a}\nonumber\\
&&  e^{{\rm i}\Lambda_{ab} P^{0ab}_0}
e^{{\rm i}\Lambda_a P^{00a}_0}
e^{{\rm i}\Lambda P^{000}_0} \cdots \nonumber\\
&&
e^{{\rm i}\Theta_{\cal I}^a G_{1/2}^a}  e^{{\rm i}\Pi_{\cal I}L_1}
 e^{{\rm i}U_{\cal I}L_0}.
\end{eqnarray}
The third exponent in the second line is chosen to be such in order to simplify
the form of resulting differential invariants.

Strictly speaking,
the expression (\ref{group}) describes
the parameterizations of the coset space of the diffeomorphisms
group over its $SO(N)$ subgroup generated by the rotations
in the odd subspace of the proper-time superspace\footnote{The parameterizations
like \p{group} was firstly introduced in \cite{IK1,IK2} for 2-dimensional
(super)conformal groups.}. Since the quantities
which will be used for the construction
of the action are inert with respect to the  stability subgroup,
one can consistently restrict the consideration to this coset space.

The elements $G_{\cal I}$ (\ref{group}) differ
from each other only in the last line, which represents
the subgroup, generated by dilation $L_0=P^0_0+1/2 P^a_a$,
one dimensional conformal
boost $L_1=P^{00}_0+ P^{0a}_a$ and $N$ superconformal
transformations $G_{1/2}^a=-i P^{0}_a+P^{0a}_0+P^{ab}_b$.
So, all $(D+2)$ elements have the same parameters in the group space,
except the last three
ones $ \Theta_{\cal I}^a, \Pi_{\cal I}, U_{\cal I}$.
This property of the elements is unchanged after the left action
\begin{equation}\label{left}
G'_{\cal I}=G_0 G_{\cal I}
\end{equation}
 with  an arbitrary constant group element $G_0$. It is a consequence of the
 representation
 \begin{equation}\label{kh}
G_{\cal I}=K H_{\cal I}
\end{equation}
in which parameters of the coset $K$ transform independently from the
parameters of the subgroup $H_{\cal I}$\cite{CWZ}-\cite{PK}.
In spite of the fact that the element $G_0$ is a constant element, it contains
infinite number of parameters, similarly to the elements (\ref{group}).
Under such (infinitesimal) transformations
\begin{equation}\label{inf}
G_0=1+i\epsilon
\end{equation}
the first
two parameters $\tau$ and
$\vartheta^a$ transform exactly
as the coordinates $x^M=\{\tau,\vartheta^a\}$ of the $(1,N)$  su\-per\-space\cite{P_0}
\begin{equation}\label{trans}
\delta \tau = \epsilon(\tau,\vartheta^b),\quad
\delta \vartheta^a = \epsilon^a(\tau,\vartheta^b),
\end{equation}
where infinitesimal superfunctions $\epsilon(\tau,\vartheta^b)=\epsilon^0(x^M)$ and
$\epsilon^a(\tau,\vartheta^b)=\epsilon^a(x^M)$ are constructed out of parameters
of the infinitesimal  element
\begin{equation}
\epsilon =\epsilon^M P_M+{\epsilon^{M}}_{M_1}{P^{M_1}}_M+
{\epsilon^{M}}_{M_1M_2}{P^{M_2M_1}}_M+... .
\end{equation}
 which belong to the
superdiffeomorphisms algebra. The explicit form of the $\epsilon^M(x)$
is the following
\begin{equation}
\epsilon^M(x)=\epsilon^M+{\epsilon^M}_{M_1}x^{M_1}+
{\epsilon^M}_{M_1M_2}x^{M_2}x^{M_1}+...\;.
\end{equation}

One can consistently consider all other parameters in the group element
as superfields -
 as functions of all
these superspace coordinates. As it was already mentioned, such superfield
approach
faces  difficulties when $N\geq 3$.  So, in this paper
we will consider the so called spontaneously broken
realization, i.e., when the parameters $\vartheta^a$ instead of being
the Grassmann coordinates
of the superspace are the Goldstone fields
$\vartheta^a(\tau)$ which depend on  only bosonic coordinate
$\tau$. All other parameters are the functions of $\tau$ as well,
though we do not write this explicitly for shortness. The transformation laws
\p{trans} look in this case as
\begin{equation}\label{trans1}
\delta \tau = \epsilon(\tau,\vartheta^b(\tau)),\quad
\delta \vartheta^a = \epsilon^a(\tau,\vartheta^b(\tau)).
\end{equation}
\section{Construction of the Invariant Action}
As is well known, the Cartan's differential forms $\Omega_{\cal I}=
G_{\cal I}^{-1} d G_{\cal I}$ are invariant under the transformations
(\ref{left}). Their expansion coefficients at different generators
in the series
\begin{equation}\label{Omega}
\Omega_{\cal I}=i{\omega_{\cal I}}^0 P_0+i{\omega_{\cal I}}^a P_a+\ldots+
i{\omega_{\cal I}}^0_{00} P^{00}_0+\ldots
\end{equation}
are invariant differential one-forms which can be used
for the construction of invariant action integrals.
In one-dimensional space which is under the consideration
the external products of one-forms vanish and
the only nonvanishing differential invariants are linear
combinations of the coefficients $\omega_{\cal I}$.

Consider the following expression for the action
\begin{equation}\label{GenAct}
  S=-\frac{1}{2}\Sigma_{\cal I}\int {\omega_{\cal I}}^0_{00},
\end{equation}
where $\Sigma_{\cal I}=(-++\ldots++-)$
is the signature of $D+2$ - dimensional
space-time and summation over external index ${\cal I}$ is assumed.

The action (\ref{GenAct}) exactly coincides with the action
for $N=1$ spinning particle\cite{P2} and is
analogous to the Trautman's formulation\cite{T} of the gravity
(in the form which admits the one-dimensional
consideration\cite{P_0}). The expression for the omega-form
${\omega_{\cal I}}^0_{00}$ is rather complicated and
contains all parameters explicitly written in the expression (\ref{group}).
To make the connection with the known action for the spinning
particle transparent, one have to  gauge away some of them. Indeed, the
transformation law (\ref{trans}) shows that the Goldstone fields
$\vartheta^a$ can be  vanished by the appropriate choice of the
parameters $\epsilon^a(\tau,\vartheta^b)$. Moreover, the
residual symmetry can be used to exclude some additional fields\cite{P_0},
namely the fields $\rho_a, {\cal C}_{ab}$ and $\zeta_{ab}^c$.
Indeed, the transformation laws of all these fields contain terms,
proportional to the derivatives of the transformation parameters
$\epsilon(\tau,\vartheta^b(\tau))$ and $
\epsilon^a(\tau,\vartheta^b(\tau))$\cite{P_0}:
\begin{eqnarray}\label{highTrans}
\delta\rho_a(\tau)&=&\frac{\delta}{\delta\vartheta^a(\tau)}
\epsilon(\tau,\vartheta^d(\tau))+\cdots ,\\
\delta C_{ab}(\tau)&=&\frac{\delta^2}
{\delta\vartheta^a(\tau)\delta\vartheta^b(\tau)}
\epsilon(\tau,\vartheta^d(\tau))+\cdots ,\\
\delta\zeta_{ab}^c(\tau)&=&\frac{\delta^2}
{\delta\vartheta^a(\tau)\delta\vartheta^b(\tau)}
\epsilon^c(\tau,\vartheta^d(\tau))+\cdots .
\end{eqnarray}
These derivatives are indeed the variational derivatives and they have to be calculated
at the point $\vartheta^a(\tau)=0$ because we already have made this gauge choice.

The explicit expression for the ${\omega_{\cal I}}^0_{00}$ in this gauge is
\begin{eqnarray}\label{omega000}
{\omega_{\cal I}}^0_{00}&=&d\tau e^{U_{\cal I}}\{\Pi_{\cal I}^2+{\dot \Pi_{\cal I}}-
i {\dot \Theta_{\cal I}^a}\Theta_{\cal I}^a-\\\nonumber
&&-3\Lambda-\chi_a\kappa_a+\dot\chi_a\chi_a-
i\chi_a\Lambda_a+\\\nonumber
&&2i\Lambda_a\Theta_{\cal I}^a -2\kappa_a\Theta_{\cal I}^a-
2\Lambda_{ab}\chi_b\Theta_{\cal I}^a-
2i\A_{ab}\chi_b\Theta_{\cal I}^a+\\\nonumber
&&\Lambda_{ab}\Theta_{\cal I}^a\Theta_{\cal I}^b-
i\A_{ab}\Theta_{\cal I}^a\Theta_{\cal I}^b\}.
\end{eqnarray}
The rest of parameters $\chi_a, \A_a^b$ and $\kappa^a$ are eated
by parameters $\Lambda, \Lambda_a$ and
$\Lambda_{ab}$ which become, correspondingly, $\tilde\Lambda,
\tilde\Lambda_a$ and
$\tilde\Lambda_{ab}$:
\begin{eqnarray}\label{redef}
\tilde\Lambda&=&\Lambda+\frac{1}{3}\chi_a\kappa_a-\frac{1}{3}\dot\chi_a\chi_a+
\frac{i}{3}\chi_a\Lambda_a,\\\nonumber
\tilde\Lambda_a&=&\Lambda_a +i\kappa_a+
i\Lambda_{ab}\chi_b-
\A_{ab}\chi_b,\\\nonumber
\tilde\Lambda_{ab}&=&\Lambda_{ab}-
i\A_{ab}.
\end{eqnarray}
In addition, one can eliminate the field
$\Pi_{\cal I}$ with the help of its equation of motion
\begin{equation}
\Pi_{\cal I}=\frac{1}{2}\dot U_{\cal I}.
\end{equation}

In terms of new variables
\begin{equation}
x_{\cal I}=e^{U_{\cal I}/2},\;\;
\Psi_{\cal I}^a=e^{U_{\cal I}/2}\Theta_{\cal I}^a
\end{equation}
the action (\ref{GenAct}) gets the familiar form\cite{S}
\begin{eqnarray}               \label{Nconf}
S&=&\frac{1}{2}\int{\rm d} \tau(\dot{x}^2_{\cal I}+
i\dot\Psi_{\cal I}^a\Psi_{\cal I}^a+\\\nonumber
&&3\tilde\Lambda x^2_{\cal I}-2i\tilde\Lambda_a\Psi_{\cal I}^ax_{\cal I}-
i\tilde\Lambda_{ab}\Psi_{\cal I}^a\Psi_{\cal I}^b).
\end{eqnarray}
So, the action (\ref{GenAct}) which is
invariant with respect to the diffeomorphisms group
of the proper-time superspace $(1,N)$ properly describes
the $N$ - extended spinning particle.
\section{Conclusions}
In the framework of nonlinear realizations of infinite - dimensional
diffeomorphisms groups of  the $(1,N)$
superspace we have constructed the reparametrization
invariant actions
for $N$-extended spinning particle in arbitrary
dimension $D$. It is achieved by simultaneous consideration of
$D+2$ group elements. The parameters of corresponding group points
include the coordinates and momenta of the particle.
The interaction between coordinates which effectively reduces the number of
the space-time dimensions from $D+2$ to $D$ is included in the action
by the presence of the parameters
with higher dimensions, which play the role of the Lagrange multipliers
and are  the same for all considered $D+2$
points on the group space.

It is worth mentioning, that the action obtained for the $N$- extended
spinning particle have the same form as the Trautman's\cite{T} action
for the gravity\cite{P_0}.  The first of them is invariant under the
reparametrizations of  the $(1,N)$
superspace, having only one bosonic and $N$ Grassmann coordinates. The second one
is written down in $D$-dimensional bosonic space-time. This analogy is very intriguing.
So, it would be interesting to apply the method developed here
to  higher dimensional superspaces to construct corresponding
supergravity theories. In particular, such approach in two-dimensional
(super)space can be useful to construct the Marnelius - like description
of the bosonic strings and superstrings. The another possibility is the
analogous consideration of the nonlinearly realized W-algebras giving the
symmetries of the particle with rigidity.

\section*{Acknowledgments}
A.P. would like to thank Prof. S. Catto for the kind hospitality in the
City University of New York and Profs. P. Pasti, D. Sorokin and M. Tonin
for valuable discussions and for the kind hospitality in the
University of Padova, where the
essential part of this work was done.
The work  was supported in part by the
INTAS grant 00-00254.

\end{document}